\documentclass{PoS}
\usepackage{graphicx}
\usepackage{graphics,epsfig}
\PoS{PoS(BDMH2004)057}

\title{High resolution SPH simulations of disk formation in CDM halos;
resolution tests}

\ShortTitle{Numerical influences on galaxy formation}

\author{\speaker{Tobias Kaufmann}\\
        Inst. for Theoretical Physics, Univ. of Zürich, Switzerland\\
        E-mail: \email{tkaufmann@physik.unizh.ch}}

\author{{Lucio Mayer$^{1}$, Ben Moore$^{1}$, Joachim Stadel$^{1}$ \& James Wadsley$^{2}$}\\
        $^{1}$Inst. for Theoretical Physics, Univ. of Zürich, Switzerland\\
  $^2$Department of Physics \& Astronomy, McMaster University, Canada}

\abstract{We perform N-Body/SPH simulations of
disk galaxy formation inside equilibrium spherical and triaxial 
cuspy dark matter halos. We systematically study the disk properties and 
morphology as we increase the numbers of dark matter and gas particles from $10^4$ to $10^6$ and change the force resolution. The force resolution influences the morphological evolution of the disk quite dramatically. Unless the baryon fraction is significantly lower than the universal value, with high force 
resolution
a gaseous bar always forms within a billion years after allowing cooling to begin. The bar interacts with the disk, transferring angular momentum 
and increasing its scale length. In none of the simulations does the final mass
distribution of the baryons obey a single exponential profile. Indeed within
a few hundred parsecs to a kiloparsec from the center the density rises much
more steeply than in the rest of the disk, and this is true irrespective of
the presence of the bar.}

\FullConference{Baryons in Dark Matter Halos\\
		 5-9 October 2004\\
 		 Novigrad, Croatia}

\begin{document}
\section{Introduction and numerial models}

Although there is a quite extensive literature on the analysis of numerical
effects in cosmological SPH simulations of structure formation 
(e.g. Steinmetz \& White 1997), often
the complexity of such simulations makes it nearly impossible to clearly
separate what is physical from what is numerical. Attempts have been
made to try single out various numerical effects by running simple
tests on rotating spheres of gas and dark matter and trying to extrapolate
results to the case of the full cosmological simulations (Okamoto
et al. 2003). However, these tests adopt quite idealised initial conditions
that are barely related to the conditions of cosmic structures in a CDM model. 
Moreover, even with such idealised experiments, current 
resolution tests do not probe a large enough range of particle 
numbers or force resolution to demonstrate convergence. However, 
we wish to understand if we can trust the current state of the
art galaxy formation simulations that end up of with about
$10^5$ SPH and dark matter particle in single object at 
$z=0$ (e.g. Governato et al. 2004).
We have studied numerical resolution effects that occur
during the dissipative cooling of gas within isolated equilibrium
dark matter halos. At our best resolution we use a million particles
per halo and hundred parsec force softening. The halo models have 
structural properties and shapes that resemble as closely as possible
those arising in CDM simulations. In this high resolution regime artificial heating of
gas particles  by dark matter particles
(Steinmetz \& White 1997)
%and loss of angular momentum by artificial viscosity
%(Thacker et al. 2000)
should be already under control
($N > 10^4$ is required to mitigate those effects).

%\section{Models}

In our models we follow the formation of disks inside haloes on scales comparable
to the Milky Way and M33 (respectively $V_{circ}\sim140$ and $\sim115$ km/s) and 
with different mass resolutions: LR ($3\cdot10^{4}$),
IR ($9\cdot10^{4}$ ), HR ($10^{6}$, $5\cdot10^{5}$) particles in the dark matter and the gas phases respectively.
Our initial conditions comprise in one case of a spherical CDM like halo (Kazantzidis 2004) built with structural parameters expected in the concordance
$LCDM$ model for the chosen mass scale, and an
embedded hot gaseous halo in (approximate) hydrostatic equilibrium. 
Both components are spinning and have the
same specific angular momentum distribution, which we
take from results of cosmological simulations (Bullock et al. 2001).
The second set of initial conditions we create by merging non-rotating gaseous
plus dark matter halos on different orbits (see Moore et al. 2004).
These produce equilibrium triaxial or axisymmetric prolate and oblate 
structures with spin parameters in the same range as found within halos from
hierarchical simulations. In these paper we will show the results from runs
using the first type of initial conditions.
In our highest resolution runs we resolve
the gas phase down to $10^{5}M_{\odot}$ or $2\cdot10^{-5}\%$ of $M_{vir}$. The dark matter halos have
an NFW density distribution and spin parameter an $\lambda=J\left|E\right|^{\frac{1}{2}}/(GM^{\frac{5}{2}})$
from $0.045$ to $0.1.$ The gas contributes in mass from $6\%$ to $9\%$ of 
the total mass of the system. We use the parallel Tree+SPH code GASOLINE (Wadsley et al (2004)) 
and a standard cooling function for a primordial gas of helium and (atomic) hydrogen.

\section{Force resolution; disk scalelengths and morphologies}

In this section we will discuss the results of the Milky Way-sized models,
which have the highest baryon fraction ($9 \%$).
We ran all simulations with a spline softening of $h_{1}=0.5$ kpc and $h_{2}=2$ kpc and focus now on the effects of changing force resolution. The smaller
softening leads to the formation of a strong gaseous 
bar of length comparable to the size of the disk at that time, 
while the disk continues to grow from the inside out
with continued gas accretion. Larger values of the softening, typical
of that used within most cosmological simulations, suppress the bar.  Since  $h_{2}$ is of order of the disk scale length
one expects modes with wavelengths at such scales to be damped (with
a spline kernel as the one we adopt softening of gravity disappears only
at twice  $h_{2}$). In fact in the IR run with softening $h_{2}$  a bar still
forms but dissolves in less than 1 Gyr.
\begin{figure}
\includegraphics[scale=0.28]{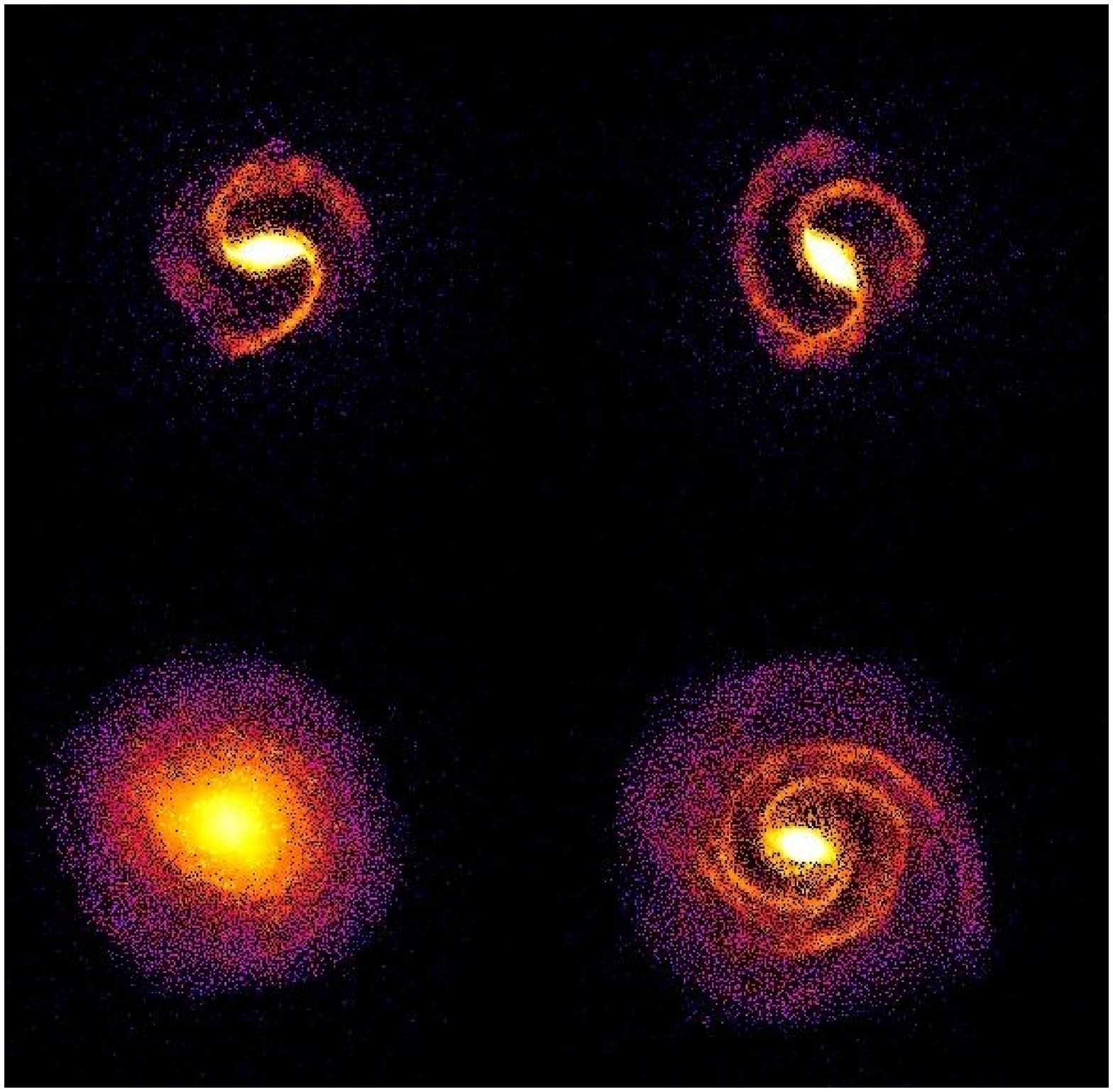} 
\includegraphics[scale=0.3]{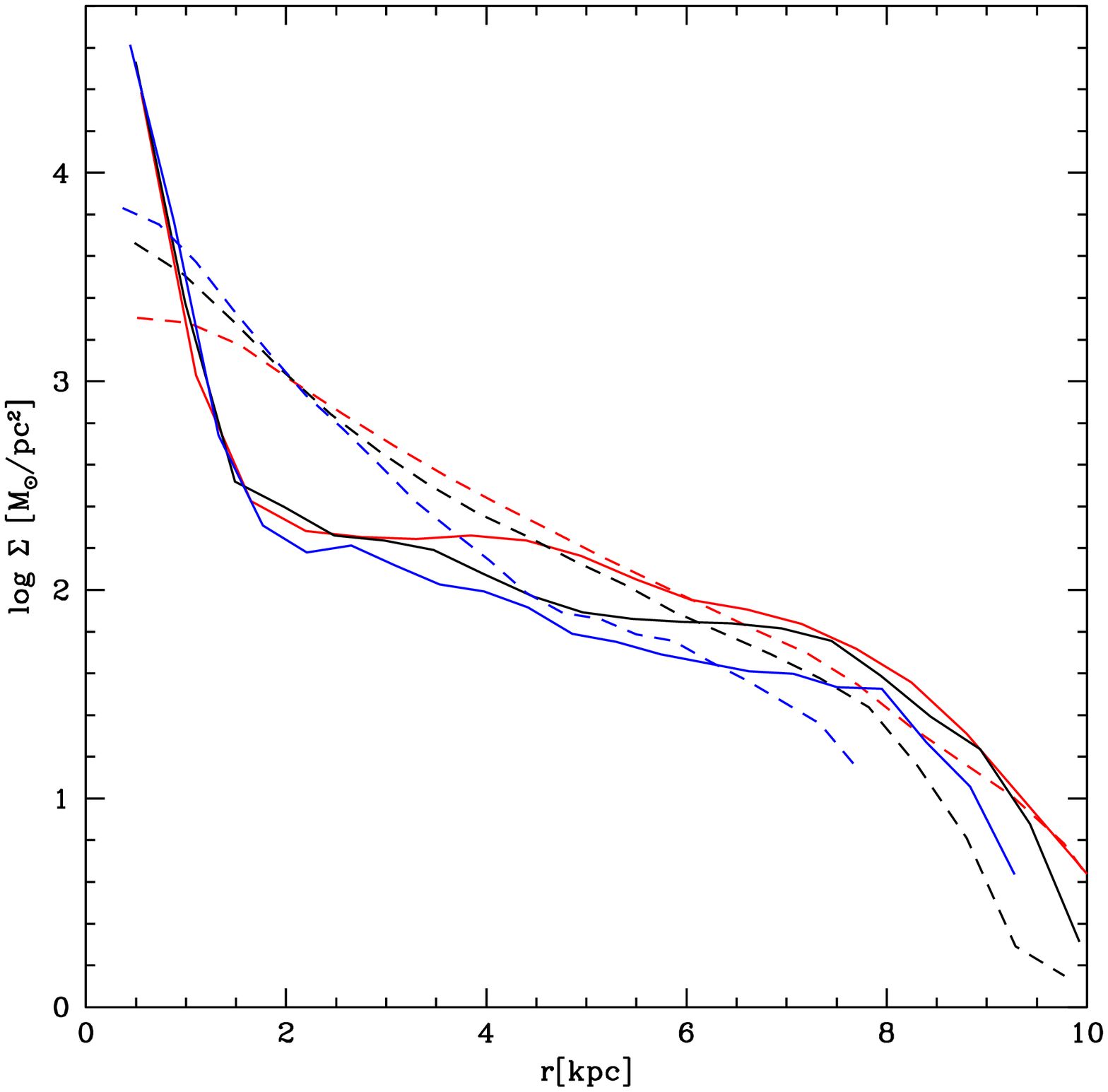}
\caption{Face on views of the gas density within models having softening 
$h_{2}$ (left) and $h_{1}$ (right), IR model. The upper and lower pictures show a snapshot after 1 Gyrs and 6 Gyrs respectively within a frame of length 
25 kpc on a side. The graph in the right panel shows the logarithmic
surface density of the gas disk after 8 Gyr. Run HR, IR and LR are plotted in red, black and blue
respectively. Dashed/solid lines: softening 2 kpc/0.5 kpc.
}

\end{figure}

\begin{figure}
\includegraphics[scale=0.3]{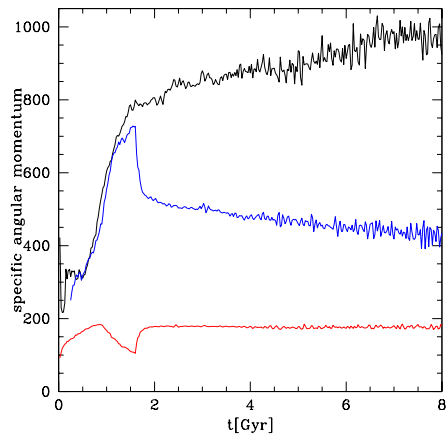}

\caption{Influence of the bar to the evolution of angular momentum of gas disk particles. We followed a ring of particles which never reach the inner two kpc,
hence they always stay outside the bar: IR model, in black:  softening 0.5 kpc, in blue: softening 2 kpc, in red:  angular momentum of the particles which were initially in the bar, softening 2 kpc. How much the angular momentum of such particles grows
appears to be correlated with the formation of the bar:
the bar dissolves for the big softening shortly before 2 Gyr.
\label{cap:plot angmomevo}
}
\end{figure}

Assuming conservation of angular momentum and 
an exponential surface density profile for the cold disk material
Mo, Mao \& White (1998, MMW) calculated the disk scalelength to be
$R_{d}=\frac{1}{\sqrt{2}}\left(\frac{j_{d}}{m_{d}}\right)\lambda r_{200}f_{c}^{-0.5}f_{R}(\lambda,c,m_{d},j_{d})$. The expected scale length for the
Milky Way simulations should be $R_{d}=2.2$ kpc  
for $\lambda=0.045$ and by replacing $r_{200}$ with the cooling radius (which is $110$ kpc in HR model) in the simulations. For the HR models we find scalelengths of 
$\sim 2$ or $\sim 3$ kpc depending on whether the bar is absent or
present (smaller scale lengths are found in LR and IR models), see Figure 1. 
The gaseous bar affects heavily the evolution of the baryonic 
surface density, and thus the disk scale length, by
transporting angular momentum from the inner part to the outer
part of the galaxy. This can be seen in Figure \ref{cap:plot angmomevo}, where 
the evolution of the specific angular momentum of some disk gas particles which initially formed a ring around the bar, is plotted. 
Such transport of angular momentum of course has a duration comparable 
to the lifetime of the bar, which varies considerably depending on the
softening as we noted, and is also affected by the adopted mass resolution
(number of particles). The transition from a barred galaxy to a nonbarred one 
in the IR model is apparent in Figure 1 and 2.
 
An analytic estimate of the stability
against bar formation is given by the $\epsilon_{d}$ parameter (
Efstathiou, Lake \& Negroponte 1983, Mayer \&Wadsley 2004): $\epsilon_{
d}=V_{peak}/\sqrt{(GM_{disk}/R_{d}},$ where $V_{peak}$ is the peak rotational velocity usually reached
around $2R_{d}$ and $M_{disk}$ is the total mass of the disk. We
measured $\epsilon_{d}$ for one of our barred simulations at 0.75
Gyr - shortly before it went bar unstable - and got $\epsilon_{d}=0.60,
$
which is clearly smaller than $0.94,$, the value found to be required
for stability of gas dominated disks
against bar formation (see Mayer \&Wadsley 2004). 
%We will discuss the effect of bars and their evolution in a forthcoming paper.

\section{Initial baryon fraction and spin parameter}

The baryon fraction plays an important role. Decreasing the initial
baryon fraction to $6\%$ and using $\lambda=0.1$ lowers the final amount of cold gas by $47\%$ and completely changes the morphological evolution of the disk.
Even with high force resolution (softening $h=250$pc) the disk remains
stable and no bar develops, although we 
found a small nucleus in the centre and some very weak spiral arms.

\begin{figure} \includegraphics[scale=0.32]{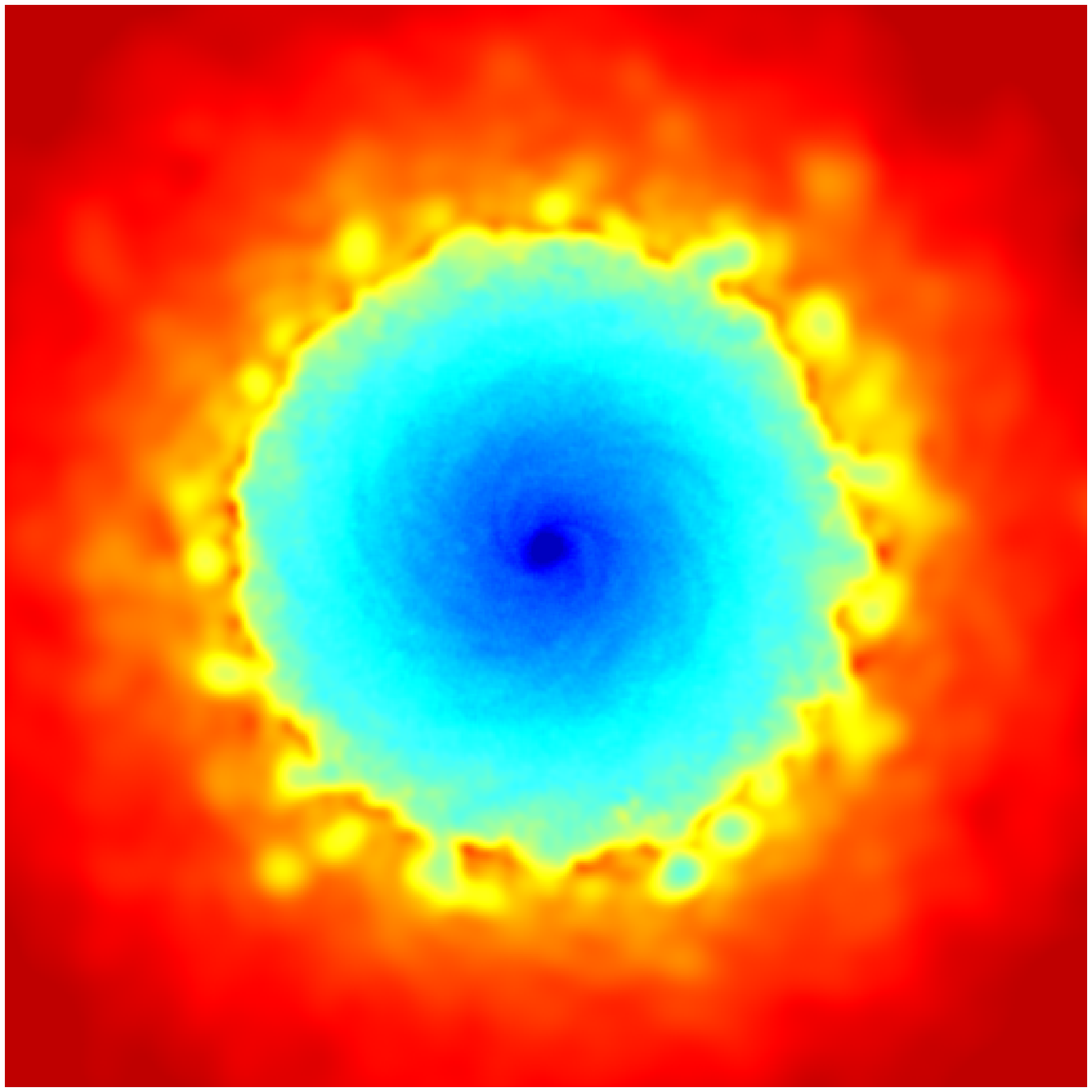} \includegraphics[scale=0.3]{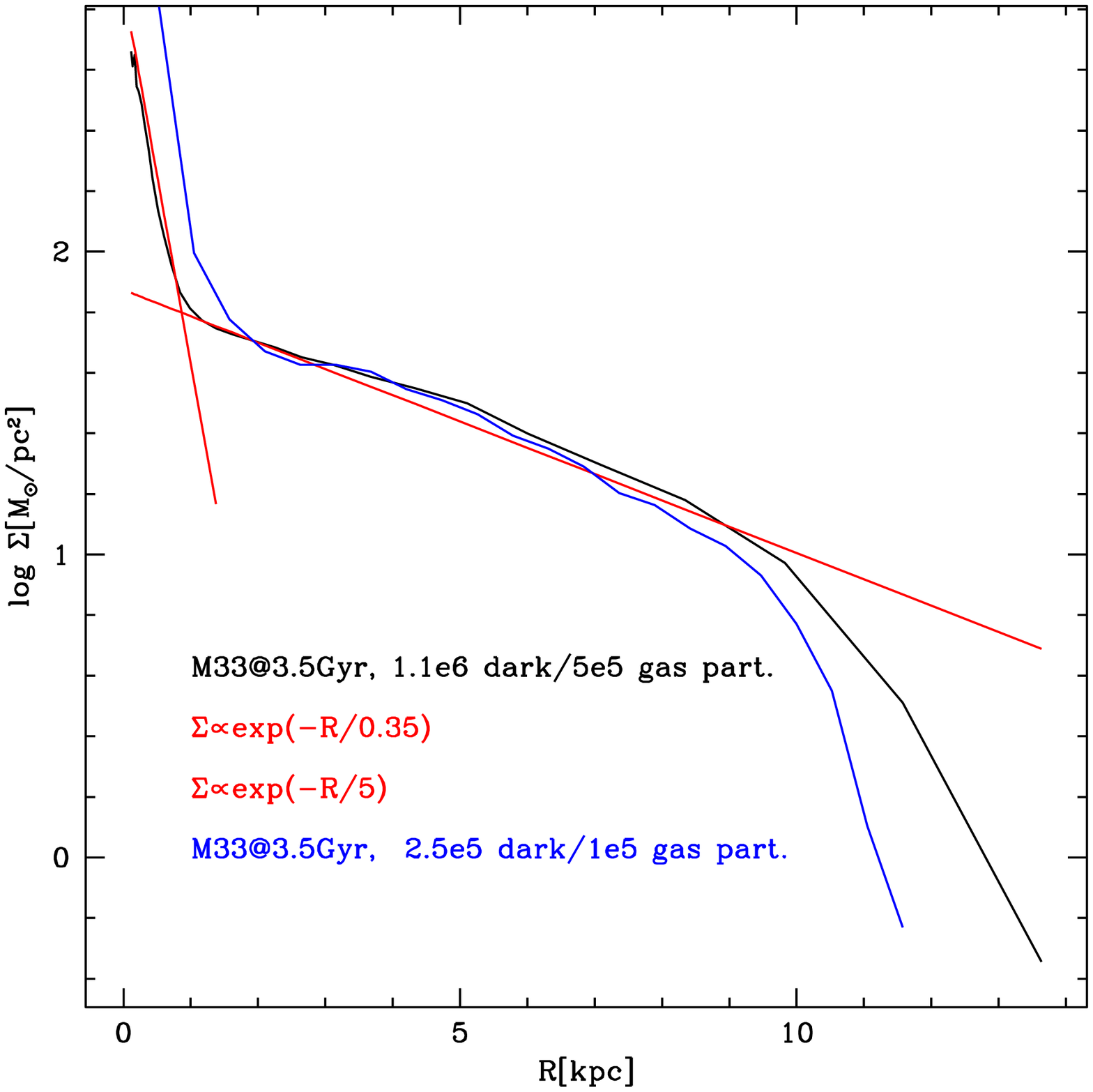}

\caption{The left panel shows a colour density map of the high resolution 
M33 gasdisk after 3.5 Gyr. The transient spiral patterns closely resemble
Sc-Sd galaxies. The right panel shows the surface density profile with fitting curves of this model.}
\end{figure}
This high resolution galaxy model provides an ideal testbed for 
the analytic model of Van den Bosch (2001) for disk structure.
Assuming cuspy CDM halos and initial gaseous angular momentum
profiles similar to the dark matter, he finds that 
it is impossible to form a pure exponential disk, in all cases
a central luminosity spike results from central low angular momentum
material.
Indeed we are also unable to find a pure exponential disk and we find 
a steeper profile in the center in our simulations (Figure 3).
However our M33 model 
only has a small nuclear gas spike and it is possible that with
even higher resolution this material collapses to form a dense
nucluear star cluster as observed within galaxies such as M33.
It remains to be seen whether the mass distribution is comparable
to that in the models by Van den Bosch (2001), which also included
star formation but do not fully include disk self-gravity. 
%(the latter
%might be responsible for angular momentum transport and inward mass flux). 

We resolve for the first time the mass better than $~10^{5}M_{\odot}$.
At such resolution there is clumping of infalling gas visible at the edge
of the disk; this is the first time that such clumpy gas accretion is
witnessed in numerical simulations.

\section{Summary}

We simulated the formation of a gaseous disk embedded in a dark matter
halo at very high resolution.  Bar formation is a very common process - particularly with high baryon fractions, low spin parameters, and needs high force resolution. Independent of the adopted temperature floor all the ``Milky Way''
disks become bar unstable as soon as we reached a force resolution of $0.5$ kpc
or $\sim0.25\%$ of $r_{vir}$. This picture changes in the M33 model with lower mass and baryon fraction: An exponential outer disk plus a small dense exponential nucleus are formed in this case.\\

We would like to thank Stelios Kazantzidis for providing a code to generate isolated dark matter halos.

\end{document}